\documentclass[12pt,english]{article}
\pdfoutput=1
\usepackage{graphicx,array}
\usepackage{cite}
\usepackage{color}
\usepackage{latexsym}
\usepackage{amsthm}
\usepackage{amsmath}
\usepackage[titletoc]{appendix}
\usepackage{enumitem}
\usepackage{amssymb}
\usepackage[unicode=true,
 bookmarks=true,bookmarksnumbered=false,bookmarksopen=false,
 breaklinks=false,pdfborder={0 0 1},backref=false,colorlinks=true]
 {hyperref}
\usepackage{breakurl}
\usepackage[hang,flushmargin]{footmisc} 
\usepackage{natbib}

\def\simgt{\mathrel{\lower2.5pt\vbox{\lineskip=0pt\baselineskip=0pt
           \hbox{$>$}\hbox{$\sim$}}}}
\def\simlt{\mathrel{\lower2.5pt\vbox{\lineskip=0pt\baselineskip=0pt
           \hbox{$<$}\hbox{$\sim$}}}}

\newcommand{\be}{\begin{equation}}
\newcommand{\ee}{\end{equation}}
\newcommand{\bea}{\begin{eqnarray}}
\newcommand{\eea}{\end{eqnarray}}

\newcommand{\ham}{\hat{H}}

\newcommand{\HH}{\mathcal{H}}

\newcommand{\ket}[1]{\left| #1 \right\rangle}
\newcommand{\bra}[1]{\left\langle #1 \right |}

\definecolor{nicered}{rgb}{0.7,0.1,0.1}
\definecolor{nicegreen}{rgb}{0.1,0.5,0.1}
\hypersetup{colorlinks,citecolor=black,linkcolor=black,urlcolor=black}

\begin{document}
\baselineskip=14pt
\hfill CALT-TH-2021-010
\hfill

\vspace{2cm}
\thispagestyle{empty}
\begin{center}
{\LARGE\bf
Reality as a Vector in Hilbert Space
}\\
\bigskip\vspace{1cm}{
{\large Sean M.\ Carroll}\footnote{e-mail: \url{seancarroll@gmail.com}}
} \\[7mm]
 {\it Walter Burke Institute for Theoretical Physics\\
    California Institute of Technology,
   Pasadena, CA 91125\\
   and Santa Fe Institute, Santa Fe, NM 87501} \\
 \end{center}
\bigskip
\centerline{\large\bf Abstract}

\begin{quote} \small
I defend the extremist position that the fundamental ontology of the world consists of a vector in Hilbert space evolving according to the Schr\"odinger equation.
The laws of physics are determined solely by the energy eigenspectrum of the Hamiltonian.
The structure of our observed world, including space and fields living within it, should arise as a higher-level emergent description.
I sketch how this might come about, although much work remains to be done.

Invited contribution to the volume \emph{Quantum Mechanics and Fundamentality: Naturalizing Quantum Theory Between Scientific Realism and Ontological Indeterminacy}; Valia Allori (ed.).
\end{quote}

\newpage
\baselineskip=16pt
	
\setcounter{footnote}{0}

\begin{quote}
\emph{[I]t would be naive in the extreme to be a `Hilbert-space-vector realist': to reify Hilbert space, and take it as analogous to physical space.} 
-- \citet{wallace2017}
\end{quote}

\begin{quote}
\emph{In Hilbert space, nobody can hear you scream.} -- \citet{aharonov2005quantum}
\end{quote}


Almost a century after the 1927 Solvay Conference, the question of the ultimate ontology of quantum mechanics remains unsettled.
Essentially all formulations of quantum theory rely on the use of a wave function or state vector (or mathematically equivalent structures).
But researchers do not agree on whether the state vector is a complete and exact representation of reality, whether it represents part of reality but needs to be augmented by additional variables to be complete, or whether it is an epistemic tool rather than a representation of reality at all.
And they further do not agree on whether the state vector should be thought of as purely an element of some abstract Hilbert space, or whether there is some fundamental ontological status to a particular representation of that vector in terms of something more directly physical, such as configuration space or particles or fields in honest three-dimensional ``space.''

Here I want to argue for the plausibility of an extreme position among these alternatives, that the fundamental ontology of the world is completely and exactly represented by a vector in an abstract Hilbert space, evolving in time according to unitary Schr\"odinger dynamics.
Everything else, from particles and fields to space itself, is rightly thought of as emergent from that austere set of ingredients. 
This approach has been called ``Mad-Dog Everettianism" \citep{Carroll:2018rhc} although ``Hilbert Space Fundamentalism" would be equally accurate.

Let's see how one might end up seduced by an ideology that is so radically different from our direct experience of the world.
When we are first taught quantum mechanics, we are shown how to construct quantum theories by taking classical models and quantizing them.
Imagine we have a classical precursor theory defined on some phase space, expressed mathematically as a symplectic manifold $\Gamma$, with evolution determined by some Hamiltonian function $H: \Gamma \rightarrow \mathbb{R}$.
We choose a ``polarization'' on phase space, which amounts to coordinatizing it in terms of canonical coordinates $Q$ (defining ``configuration space'') and corresponding canonical momenta $P$, where each symbol might stand for multiple dimensions.
This is a fairly general setup; for $N$ point particles moving in $D$-dimensional Euclidean space, configuration space is isomorphic to $\mathbb{R}^{DN}$, but we could also consider field theory, for which the coordinates are simply the values of the fields throughout space.

One way to construct a corresponding quantum theory is to introduce complex-valued wave functions of the coordinates alone, $\Psi(Q) \in \mathbb{C}$.
Wave functions must be normalizable, in the sense that they are square-integrable, $\int \Psi^*\Psi \,dQ <\infty$, where $\Psi^*$ is the complex conjugate of $\Psi$.
Momenta are now represented by linear operators $\hat{P}$, whose form can be derived from the canonical commutation relations $[\hat{Q},\hat{P}]=i\hbar$ (where the operator $\hat{Q}$ is simply multiplication by $Q$).
This lets us promote the classical Hamiltonian to a self-adjoint operator $\hat{H}(\hat{Q}, \hat{P})$ (up to potential operator-ordering ambiguities).
We then posit that the wave function evolves according to the Schr\"odinger equation,
\be
  \hat{H}\Psi = i\hbar \frac{\partial}{\partial t}\Psi.
\ee
This form of the Schr\"odinger equation is perfectly general, and applies to relativistic theories as well as non-relativistic ones, as long as one uses an appropriate Hamiltonian.

This procedure gives us the beginnings of a quantum theory.
For Everettians, it gives us the complete theory; a unitarily-evolving quantum state describes the entire ontology at a fundamental level.
Other approaches require additional dynamical rules, physical structures, or some combination thereof.
Here our interest is in seeing how far we can get from a minimal starting point, so the Everettian approach is appropriate.
[For more on structure in the Everett interpretation, see \citep{wallace2003everett,pittphilsci18772}.]

But how precisely should we think about this ontology?
It might seem to include at least original configuration space coordinatized by $\{Q\}$, the set of wave functions $\Psi(Q)$, and the Hamiltonian $\hat{H}$.
This has led \citet{albert1996} to suggest ``wave function realism'' -- the idea that it is the wave function $\Psi(Q)$ that describes reality, and that the wave function is defined on configuration space, so that configuration space  ($3N$ dimensional for $N$ particles in 3 dimensions, infinite-dimensional for a field theory) is where the wave function ``really lives" [see also \citep{ney2013,north2013}].
This point of view has been criticized by \citet{wallace2017}, who points out that configuration space is simply one possible way of \emph{representing} the wave function; it could also be represented in momentum space (as a function of $P$), or an infinite number of other choices.
[Wallace also has other criticisms, and other authors criticize the idea for other reasons \citep{allori2013,Myrvold:2015vly}.]

The ambiguity between defining wave functions in position versus momentum space is an example of a broader issue.
``Quantization'' does not give us a one-to-one map from classical theories to quantum ones, or even a well-defined map at all.
A single classical precursor may correspond to multiple quantum theories, due to operator-ordering ambiguities.
Moreover, distinct classical models may have identical quantizations, as in the dualities of quantum field theory.
This makes it difficult to uniquely pinpoint what a quantum theory is supposed to be a theory ``of.''
In the duality between the massive Thirring model and sine-Gordon theory, a single quantum theory can arise from classical precursors describing either fundamental bosons or fundamental fermions \citep{PhysRevD.11.2088}.
In the AdS/CFT correspondence, a nongravitational field theory in $D$ spacetime dimensions is dual to a quantum gravity theory in $D+1$ dimensions \citep{Maldacena:1997re}.
So from a quantum theory alone, it might be impossible to say what kinds of fields the model describes, or even the number of dimensions they live in.

But the complicated relationship between classical and quantum theories is a problem for physicists, not for physics.
Nature simply is quantum from the start, and the classical world arises as an emergent approximation in some appropriate limit.
If our interest is in fundamental ontology, rather than focusing on quantum theories derived by quantizing classical precursors, it would make sense to consider the inverse problem: given a quantum theory, what kind of classical limits might arise within it?

So let us think carefully about what it means to be given a quantum theory (at least from an Everettian perspective).
Wave functions, as von Neumann noted long ago, can be added and scaled by complex numbers, and have a natural inner product defined by $(\Psi,\Phi) = \int \Psi^*\Phi$.
They therefore describe a complex, normed vector space, called Hilbert space, and the fact that the Schr\"odinger equation is linear means that it respects this structure.
In vector-space language, the choice between expressing the wave function in configuration space or momentum space is simply a change of basis in Hilbert space, which presumably has no physical importance whatsoever.
The physical quantum state -- as distinguished from its representation in some particular basis -- is simply a vector in Hilbert space, sometimes called the ``state vector,'' and written in Dirac notation as $\ket{\Psi}$.

It would therefore seem natural, if our goal is to take a nature's-eye view of things and specify the correct quantum theory of the world in its own right, to define that theory as a set of state vectors $\ket{\Psi}$ in a Hilbert space $\HH$, evolving under the Schr\"odinger equation via a specified Hamiltonian $\ham$.
The problem is that this is very little structure indeed.
Hilbert space itself is featureless; a particular choice of Hilbert space is completely specified by its dimension $d = \mathrm{dim}\,\HH$.
A vector in Hilbert space contains no direct specification of what the physical content of such a state is supposed to be; there is no mention of space, configuration space, particles, fields, or any such familiar notions.
Presumably all of that is going to have to somehow emerge from the dynamics, which seems like a tall order.

One might imagine that we can somehow read off the physical structure being described from the explicit form of the Hamiltonian.
For example, if we were handed
\be
  \ham = -\frac{\hbar^{2}}{2 m} \frac{\partial^{2}}{\partial x^{2}}+\frac{1}{2} m \omega^{2} \hat{x}^{2},
\ee
we would quickly surmise that we had a simple harmonic oscillator on our hands.
But that's only because we have conveniently been given the Hamiltonian in a particularly useful basis, in this case the position basis $\ket{x}$.
That is not part of the specification of the theory itself. 
Hilbert space does not come equipped with a preferred basis; we should be able to deduce that the position basis (or some other one) is useful to our analysis of the system, rather than assuming we have been given it from the start.

Given that the only data that comes along with Hilbert space is its dimension $d$, the only other thing we have to work with is the Hamiltonian considered as an abstract operator.
That does pick out one particular basis: that of the energy eigenstates, vectors satisfying $\ham \ket{n} = E_n\ket{n}$.
(For simplicity we assume non-degenerate energy eigenvalues, so that the states $\{\ket{n}\}$ define a unique basis.)
There is no information contained in the specification of the energy eigenstates themselves; they are just a collection of orthonormal vectors.
The information about the Hamiltonian is contained entirely in its set of eigenvalues $\{E_n\}$, called the ``spectrum'' of the Hamiltonian.
A specification of a quantum theory consists entirely, therefore, of this list of real numbers, the energy eigenspectrum.

There is an important caveat to this statement.
If Hilbert space is non-separable (infinite non-countable dimension), there can be unitarily inequivalent representations of the canonical commutation relations \citep{haag55}.
It is therefore necessary to give additional information to define the theory; typically this would amount to specifying an algebra of observables.

We have good reason, however, to expect that the Hilbert space for the real world is finite-dimensional, at least if we restrict our attention to our observable universe or any other finite region of space \citep{Bekenstein:1980jp,bousso1999,Banks:2000fe,Jacobson:2012yt,Bao:2017rnv}.
The reason is gravity.
In the presence of gravity, to make a long story short, the highest-entropy configuration we can construct in a spherical region \textbf{R} of radius $r$ is a black hole, and black holes have a finite entropy $S_\mathrm{BH} = A/4G = \pi r^2/G$, where $A$ is the area of the event horizon and $G$ is Newton's constant (and we set $\hbar=c=1$).
If we decompose Hilbert space into a tensor product of a factor describing that region and one describing the rest of the world,
\be
\HH = \HH_\mathbf{R}\otimes \HH_\mathbf{E},
\label{factorization}
\ee
it follows that we have an upper bound on the dimensionality of $\HH_\mathbf{R}$, given by
\be
  \mathrm{dim}\,\HH_\mathbf{R} \leq \exp{(e^{\pi r^2/G})}.
\ee
If $\mathbf{R}$ represents our observable universe with current values of the cosmological parameters, this works out to approximately $ \mathrm{dim}\,\HH_\mathbf{R} \leq e^{e^{123}}$, which is large but still smaller than infinity.\footnote{There are a number of nuances here. In gauge theories, we cannot precisely decompose Hilbert space into factors representing regions of space. And we are somewhat cheating by invoking ``regions of space'' at all, although this will be a sensible notion on individual semiclassical branches of the universal quantum state. For elaboration see \citep{Bao:2017rnv}.}

We don't know whether the dimension of the full Hilbert space $\HH$ is finite or infinite, other than the indirect consideration that infinite time evolution in a finite-dimensional Hilbert space would lead to a proliferation of Boltzmann Brains \citep{Dyson2002,Carroll:2008yd}.
But for our current purposes, it suffices that operations confined to our observable universe effectively act on a finite-dimensional part of Hilbert space. 
In that case, the set of in-principle observables is simply all Hermitian operators on $\HH_\mathbf{R}$.
What can be observed in practice will depend on how we split further split Hilbert space into systems and observers, but the theory is completely specified by the Hilbert space and Hamiltonian, or in other words, by the discrete list of numbers $\{E_n\}$ in the energy spectrum.
By which we mean the spectrum of the Hamiltonian acting on $\HH_\mathbf{R}$ itself, assuming that interactions with degrees of freedom in $\HH_\mathbf{E}$ can be neglected for practical purposes.
Henceforth we will speak as if the finite-dimensional factor $\HH_\mathbf{R}$ is the relevant part of Hilbert space, and our world can be described by a unitarily-evolving state within it.
Technically it is more likely to be a mixed state described by a density operator, but that can always be purified by adding a finite-dimensional auxiliary factor to Hilbert space, so we won't worry about such details.

The challenge facing such an approach should be clear.
The world of our experience doesn't \emph{seem like} a vector in Hilbert space, evolving according to a list of energy eigenvalues.
It seems like there is space, and objects located in space, and those objects interact with each other, and so forth.
How in the world is all of that supposed to come from a description as abstract and featureless as a vector evolving through Hilbert space?
To make matters seemingly worse, the actual evolution is pretty trivial; in the energy eigenbasis, an exact solution to the Schr\"odinger equation from an initial state $\ket{\Psi(0)} = \sum_n \psi_n\ket{n}$ is
\be
  \ket{\Psi(t)} = \sum_n \psi_n e^{-i E_n t}\ket{n}.
\ee
Each component simply evolves via an energy-dependent phase factor.
It seems like a long way from the complicated nonlinear dynamics of our world.
This challenge accounts for the epigraphs at the beginning of this paper.

Everettians tend to think that the right strategy for understanding the fundamental nature of reality is not necessarily to start with what the world seems like and to construct an ontology that hews as closely as possible to that.
Rather, we should start with some proposed ontology and ask what it would seem like to observers (if any such exist) described by it.
Clearly, ``observers'' are not represented directly by a vector in Hilbert space, nor is the world that they observe.
What we can instead ask is whether there could be a higher-level description, emergent from our ontology, that can successfully account for our world.
Such a description is not forced on us at the God's-eye (or Laplace's-Demon's eye) view of the world.
It would always be possible to say that reality is a vector in Hilbert space, evolving through time, and stop at that.
But that's not the only thing we're allowed to say.
The search for emergent levels is precisely the search for higher-level, non-fundamental descriptions that approximately capture some of the relevant dynamics, perhaps on the basis of incomplete information about the fundamental state.
The question is whether we \emph{can} recover the patterns and phenomena of our experience (space, objects, interactions) from the behavior of our fundamental ontology.

To get our bearings, consider the classic case of $N$ massive particles moving in three-dimensional space under the rules of classical Newtonian gravity.
The state of the system is specified by one point in a $6N$-dimensional phase space.
Yet there is an overwhelming temptation to say that the system ``really lives'' in three-dimensional space, not the $6N$-dimensional phase space.
Can we account for where that temptation comes from without postulating any \emph{a priori} metaphysical essence to three-dimensional space?

There are two features of the description as $N$ particles that make it seem more natural than that featuring a single point in phase space, even though they are mathematically equivalent.
The first is that the \emph{internal dynamics} of the system are more easily interpreted in the $N$-particle language.
For example, it is immediately clear that two particles will strongly affect each other when they are nearby and the others are relatively far away. 
This kind of partial and approximate understanding of the dynamics is transparent in the $N$-particle description, and obscured in the point-in-phase-space description.
The second is that the system \emph{looks like} $N$ particles.
That is, in the real-world analogues of this toy model, when we observe the system by interacting with it as a separate physical system ourselves, what we immediately see are $N$ particles. 
There can be multiple equivalent ways of describing the internal dynamics of a system, but the one we think of as ``natural'' or describing what ``really exists'' is often predicated on how that system interacts with the outside world. 

Similar considerations apply to familiar examples of emergence, such as treating a box of many atoms as a fluid.
In this case the two descriptions are not equivalent -- the emergent fluid description is an approximation obtained by coarse-graining -- but the same principles apply.
The internal dynamics of the particles in the box are more easily apprehended in the fluid description (fewer variables and equations are required, given some short-distance coarse-graining scale), and we can measure the fluid properties directly (using thermometers and barometers and so on), while the states of each individual atom are inaccessible to us.
Emergent structures that accurately describe the dynamics of part of a system using only information accessible within the emergent description itself have every right to be thought of as ``real,'' even if they are not ``fundamental'' \citep{dennett1991real,wallace2012emergent}.

These two considerations (internal dynamics and what we see) work in tandem: we want the information we gather by observing a system to be sufficient for us to predict its subsequent behavior.
The ontology that seems most natural for us to ascribe to a physical subsystem thus depends on how that subsystem interacts with the rest of the world \citep{Zanardi:2004zz}.
For our program of Hilbert-space fundamentalism, this suggests that we should look for emergent descriptions by considering ways to factorize $\HH$ into a tensor product of factors representing different subsystems, and ask how those subsystems interact with each other.
[For certain situations we might also consider direct-sum structures \citep{Kabernik:2019jko}.]

We are therefore interested in \emph{quantum mereology}: how to decompose the whole of Hilbert space into parts such that individual subsystems have simple internal dynamics, and those dynamics are readily observed via interactions with other subsystems, given nothing but the spectrum of the Hamiltonian [\citep{Carroll:2018rhc}; for related work see \citep{brun1999classical,Hartle:2008mv,Tegmark:2014kka,Ney2020-NEYFTW}].
This seems ambitious, but turns out to be surprisingly tractable.

Consider the most basic thing we might want to describe by such a factorization, the distinction between a ``system''  representing a macroscopic object exhibiting quasi-classical behavior and an ``environment" describing degrees of freedom that passively monitor the system and lead to decoherence.
This corresponds to expressing Hilbert space as the tensor product
\be
  \HH = \HH_S \otimes \HH_E.
\ee
Fixing the dimensions of $\HH_S$ and $\HH_E$, different factorizations are related by unitary transformations that mix the two factors together.
For any specific factorization, we automatically get a decomposition of the Hamiltonian into a self-Hamiltonian for the system, another self-Hamiltonian for the environment, and an interaction term:
\be
  \ham = \ham_S \otimes \mathbb{I}_E + \mathbb{I}_S \otimes \ham_E + \ham_\mathrm{int}.
\ee
We can now ask, within the set of all possible factorizations, which ones lead to system dynamics that can be described by approximate classical behavior in appropriate circumstances?

In this decomposition, internal dynamics of the system are governed by $\ham_S$ and how it is observed by the environment is governed by $\ham_\mathrm{int}$.
(We assume that $\mathrm{dim}\,\HH_E \gg \mathrm{dim}\,\HH_S$ and that the environment's internal dynamics are largely irrelevant.)
To recover classical behavior when the system is macroscopic, we want localized wave packets in $\HH_S$ to stay relatively localized and follow classical equations of motion under $\ham_S$, but also for unentangled states in $\HH_S\otimes \HH_E$ to remain relatively unentangled; the environment is supposed to passively monitor the system, not rapidly reach maximal entanglement with it.
It was argued by \citet{Carroll:2018rhc} that generic Hamiltonians, as defined by their spectra, have neither of these features in any factorization; the Hamiltonian of the real world is apparently non-generic, to nobody's surprise.

When the Hamiltonian and the system/environment split allow for classical behavior, the density operator for the system rapidly diagonalizes in the dynamically-preferred \emph{pointer basis} $\{\ket{\phi_n}\}$, such that the corresponding pointer states are robust under environmental monitoring \citep{Zurek:1981xq}.
The pointer basis defines a pointer observable in the system's Hilbert space, $\hat{Q}_S = \sum \ket{\phi_n}\bra{\phi_n}$.
Pointer states are the ones that appear classical, and a standard pointer observable is the position in space of a system; a macroscopically coherent cat is described by a pointer state, but a superposition of cats in different physical configurations is not \citep{Joos:1984uk,Zurek:1994zq,riedel2012}.
In the Quantum Measurement Limit, where $\ham_\mathrm{int}$ dominates over $\ham_S$, the pointer observable satisfies Zurek's commutativity criterion \citep{Zurek:1981xq},
\be
  [\ham_\mathrm{int}, \hat{Q}_S\otimes \mathbb{I}_E] \approx 0,
  \label{commutativity}
\ee
reflecting the fact that the system in a pointer state does not continually entangle with the environment.

These ideas about pointer states and decoherence are usually discussed within the context of a known factorization, but they also suggest a criterion for determining the factorization that allows for classical behavior \citep{Carroll:2020gme}.
Given $\HH$ and $\ham$, we search through all possible factorizations, and for each one we define a candidate pointer observable as the one that comes closest to achieving (\ref{commutativity}).
This candidate pointer observable embodies the idea that external observers ``see'' certain features of the system, and that those features evolve classically.
We can then calculate, from an initially localized and unentangled system state, the rate of delocalization and entanglement growth.
The correct factorization is the one that minimizes both.
Simple numerical examples verify that this criterion picks out what we usually think of as the standard system/environment split.
Again, the sought-after behavior is non-generic; it doesn't occur for random Hamiltonians, nor for random factorizations with any given Hamiltonian.

The data contained within the spectrum of $\ham$ therefore seems to provide the requisite information to pinpoint an appropriate emergent classical description, when one exists.
Thus reassured, let us be somewhat more ambitious, and train our sights on the ontology that seems to accurately describe our low-energy world \citep{Carroll:2021yum}: an effective quantum field theory that includes both gravity and the Standard Model of particle physics, defined in four-dimensional spacetime.

We can again take guidance from our experience of the real world.
In quantum field theory (modulo previously-mentioned nuances), we can think about degrees of freedom as being associated with regions of space.
If we partition space into regions indexed by $\alpha$, there is a corresponding decomposition of Hilbert space into a tensor product of local factors,
\be
  \HH = \bigotimes_\alpha \HH_\alpha.
  \label{localdecomp}
\ee
This lets us expand the Hamiltonian as a sum of self-Hamiltonians for each region, plus interactions between pairs of regions, triplets of regions, and so on:
\be
  \hat{H} = \sum_a h_a\hat{\mathcal{O}}_a^{\mathrm{(self)}} + \sum_{ab} h_{ab}\hat{\mathcal{O}}_{ab}^{\mathrm{(2-pt)}} + \sum_{abc} h_{abc}\hat{\mathcal{O}}_{abc}^{\mathrm{(3-pt)}} + \cdots,
  \label{localham}
\ee
where the $h_{\cdots}$ are numerical parameters.

A necessary requirement for the emergence of a structure recognizable as ``space,'' with local dynamics therein, is that the series (\ref{localham}) does not continue indefinitely when the factorization (\ref{localdecomp}) corresponds to local regions.
Degrees of freedom only interact with a small number of nearest neighbors, not with regions arbitrarily far away, which means that we only need interactions between a small number of Hilbert-space factors to capture the entire Hamiltonian.
Happily, this requirement is also essentially sufficient.
As shown by \citet{Cotler:2017abq}, generic Hamiltonians admit no local factorization at all, and when such a factorization exists, it is unique up to irrelevant internal transformations (and some technicalities that we won't go into here).
Therefore, the spectrum of the Hamiltonian is enough to pick out the correct notion of an emergent spatial structure when one exists.

A natural next step would be to define an emergent dynamical metric on spacetime, as in general relativity.
Here results are much less definitive, but initial auguries are promising.
[For alternative perspectives see \citep{hu2009emergent,Carlip:2012wa,huggett2013emergent,Nielsen:2014rfa,Giddings:2015lla,Raasakka:2016uyk,ney2020}.]
In the local decomposition of Hilbert space (\ref{localdecomp}), given an overall state $\ket\Psi$ we can calculate the reduced density matrix $\hat\rho_\alpha$ for each factor or collection of factors, and the corresponding von~Neumann entropy $S_\alpha = -\mathrm{Tr}\,\hat\rho \log \hat\rho$.
The amount of entanglement between two factors can be measured by the quantum mutual information,
\be
  I(\alpha:\beta) = S_\alpha + S_\beta - S_{\alpha\beta}.
\ee
In the vacuum state of a quantum field theory, we know that the entanglement between two regions decreases monotonically (exponentially in theories with massive fields, as a power law in conformal theories) with the distance between them.
We can therefore imagine turning this around, and \emph{defining} a distance metric depending inversely on the mutual information.\footnote{The resulting metric is defined on the same spacetime structure in which entanglement exists. This is in contrast to the emergence of spacetime from entanglement in the AdS/CFT context \citep{Swingle:2009,er-eprmvr,Faulkner:2013ica,er-eprms}, where entanglement is on the boundary and geometry is in the bulk. There is no inconsistency, as the procedure described here applies to weak-field situations far from any horizons, while the AdS/CFT construction extends over a cosmological spacetime where large-scale curvature is a central part of the description.}
If there exists an emergent best-fit smooth geometry, that can be uniquely determined using techniques such as classical multidimensional scaling.
Thus, both the dimensionality and geometry of space can be defined using the entanglement information contained in a state $\ket\Psi$, at least if that state is near to the vacuum \citep{Cao:2016mst}.

We can go further, and show that under certain optimistic assumptions (most notably, the eventual emergence of approximate Lorentz invariance), the resulting geometry obeys Einstein's equation of general relativity in the weak-field limit \citep{Cao:2017hrv}.
The basic idea, following \citet{jacobson95,Jacobson:2015hqa}, is to posit that perturbations in the quantum state lead to a change in the emergent area of a codimension-one surface that is proportional to the change in entanglement entropy across that surface,
\be
  \delta\mathcal A \propto \delta S.
\ee
The area perturbation, being a geometric quantity, can then be related to the Einstein tensor, while in the long-distance limit the entropy can be related to the stress-energy tensor.
Assuming emergent Lorentz symmetry, the resulting dynamical equation for a perturbation of the vacuum (representing flat spacetime) is
\be
  \delta G_{\mu\nu} \propto \delta T_{\mu\nu},
\ee
just as Einstein leads us to expect.

This is a provocative result, but one that shouldn't be over-interpreted.
The accomplishment is not that we recover specifically Einstein's equation; there aren't that many other locally Lorentz-invariant equations one could imagine for a dynamical spacetime metric.
What matters is that the necessary information required for such a description to emerge can be found in an abstract quantum state evolving according to a Hamiltonian specified purely by its spectrum, without any additional ontological assistance.
Moreover, while we can use entanglement to define \emph{an} emergent metric, we haven't shown that it is \emph{the} emergent metric, the one whose geodesics define the motion of localized test particles.
We can imagine how this might happen; as just one piece of the puzzle, the Lieb-Robinson bound in quantum information theory \citep{Lieb:1972wy} provides a natural mechanism whereby light-cone structures can emerge from quantum information theory \citep{Hamma:2008jt}.

What strictly emerges from entanglement in this story is the metric on space, not on spacetime; a reconstruction of the latter depends on a procedure for stitching together spatial surfaces evolving over time.
Since our entire analysis is based on the eigenvalues of the Hamiltonian appearing in the Schr\"odinger equation $\ham\ket\Psi = i \partial_t\ket\Psi$, we need to imagine that time itself is fundamental, rather than emergent.
The appearance of an explicit time parameter does not imply that this parameter is in some sense preferred.
Any Lorentz-invariant theory can be written in Hamiltonian form by choosing a frame; such a form will not be manifestly Lorentz invariant, but the symmetry is still there, so there is no obstacle to the emergent theory being approximately Lorentz- and diffeomorphism-invariant.
It would be natural if Lorentz invariance is indeed only approximate, since we are working in a finite-dimensional factor of Hilbert space, and there are no nontrivial representations of non-compact symmetry groups on finite-dimensional vector spaces.
This raises the intriguing possibility of potential experimental signatures of these ideas, as Lorentz invariance can be tested to high precision \cite{Liberati:2013xla}.
One might also worry about compatibility with the Wheeler-DeWitt equation of quantum gravity, which takes the form $\ham\ket\Psi = 0$.
In this case the eigenvalues of $\ham$ are seemingly irrelevant, since the world is fully described by a zero-energy eigenstate [cf. \citep{Albrecht:2007mm}].
But there is also no fundamental time evolution; that is the well-known ``problem of time'' \citep{Anderson:2010xm}.
A standard solution is to imagine that time is emergent, which amounts to writing the full Hamiltonian as
\be
  \ham = \ham_\mathrm{eff} - i\frac{d}{d\tau},
\ee
where $\tau$ is the emergent time parameter.
In that case everything we have said thus far still goes through, only using the eigenvalues of the effective Hamiltonian $\ham_\mathrm{eff}$.

In addition to spacetime, we still have to show how local quantum fields can emerge in the same sense as the spacetime metric. 
Less explicit progress has been made in reconstructing approximate quantum field theories from the spectrum of the Hamiltonian, but it's not unreasonable to hope that this task is more straightforward than reconstructing spacetime itself.
One promising route is via ``string net condensates," which have been argued to lead naturally to emergent gauge bosons and fermions \citep{Levin:2004mi}.

Nothing in this perspective implies that we should think of spacetime or quantum fields as illusory.
They are emergent, but none the less real for that.
As mentioned, we we may not be forced to invoke these concepts within our most fundamental picture, but the fact that they play a role in an emergent description is highly non-trivial.
(Most Hamiltonians admit no local decomposition, most factorizations admit no classical limit, etc.)
It is precisely this non-generic characteristic of the specific features of the world of our experience that makes it possible to contemplate uniquely defining them in terms of the austere ingredients of the deeper theory.
They should therefore be thought of as equally real as tables and chairs.

This has been an overly concise discussion of an ambitious research program (and one that may ultimately fail).
But the lesson for fundamental ontology is hopefully clear.
Thinking of the world as represented by simply a vector in Hilbert space, evolving unitarily according to the Schr\"odinger equation governed by a Hamiltonian specified only by its energy eigenvalues, seems at first hopelessly far away from the warm, welcoming, richly-structured ontology we are used to thinking about in physics.
But recognizing that the latter is plausibly a higher-level emergent description, and contemplating the possibility that the more fundamental vocabulary is the one straightforwardly suggested by our simplest construal of the rules of quantum theory, leads to a reconstruction program that appears remarkably plausible.
By taking the prospect of emergence seriously, and acknowledging that our fondness for attributing metaphysical fundamentality to the spatial arena is more a matter of convenience and convention than one of principle, it is possible to see how the basic ingredients of the world might be boiled down to a list of energy eigenvalues and the components of a vector in Hilbert space.
If it did succeed, this project would represent a triumph of unification and simplification, and is worth taking seriously for that reason alone.

\begin{center} 
 {\bf Acknowledgments}
 \end{center}

It is a pleasure to thank Ashmeet Singh, Charles Cao, Spiros Mikhalakis, and Ning Bao for collaboration on some of the work described here.
This research is funded in part by the Walter Burke Institute for Theoretical Physics at Caltech, by DOE grant DE-SC0011632, and by the Foundational Questions Institute.

\bibliographystyle{apa-good}
\bibliography{qmessayrefs}

\end{document}